# CDN Slicing over a Multi-Domain Edge Cloud

Tarik Taleb, Pantelis A. Frangoudis, Ilias Benkacem, and Adlen Ksentini

**Abstract**—We present an architecture for the provision of video Content Delivery Network (CDN) functionality as a service over a multi-domain cloud. We introduce the concept of a CDN slice, that is, a CDN service instance which is created upon a content provider's request, is autonomously managed, and spans multiple potentially heterogeneous edge cloud infrastructures. Our design is tailored to a 5G mobile network context, building on its inherent programmability, management flexibility, and the availability of cloud resources at the mobile edge level, thus close to end users. We exploit Network Functions Virtualization (NFV) and Multi-access Edge Computing (MEC) technologies, proposing a system which is aligned with the recent NFV and MEC standards. To deliver a Quality-of-Experience (QoE) optimized video service, we derive empirical models of video QoE as a function of service workload, which, coupled with multi-level service monitoring, drive our slice resource allocation and elastic management mechanisms. These management schemes feature autonomic compute resource scaling, and on-the-fly transcoding to adapt video bit-rate to the current network conditions. Their effectiveness is demonstrated via testbed experiments.

## 1 INTRODUCTION

Internet traffic is dominated by data distributed over Content Delivery Network (CDN) infrastructures, and a significant share of this traffic volume is due to video content. At the same time, with the increase in mobile data rates, video streaming is becoming one of the most popular services for mobile consumers. This trend is expected to strengthen in view of $5^{th}$ generation mobile networks (5G), which represent the next major phase of the mobile telecom industry, going beyond the current Long Term Evolution (LTE) and IMT-advanced systems. In fact, mobile video traffic, which accounted for 55% of the total mobile traffic in 2015, will represent more than 75% in 2020 [1].

5G comes with increased peak bit rates, higher spectral efficiency, better coverage, and the support of large numbers of diverse connectable devices, including machine type communications (MTC) ones. 5G systems are required to be cost-efficient, flexibly deployable, elastic, and above all programmable. The need to lower mobile infrastructure costs and render their deployment flexible and elastic has become critical for the sustainability of mobile operators worldwide, mainly due to the growing mobile data traffic on one hand and the stagnant average revenue per user (ARPU) on the other hand. With current mobile network designs, such required flexibility and elasticity are hard to realize, particularly due to the traditional usage of specific-purpose networking equipment that can neither dynamically scale with mobile traffic nor be easily upgraded with new functions.

From the perspective of a telecom operator, mobile Internet usage trends and the promise of 5G come with both challenges and opportunities: High-quality video delivery for users on-the-go with heterogeneous device and connectivity profiles and, importantly, time- and space-varying traffic demand can stress both the core and the radio access networks. At the same time, telecom operators strive to get more involved in the content delivery value chain; their inherent user proximity and network and infrastructure awareness are features that they can exploit to gain a competitive advantage, assuming a more active role in Quality-of-Experience-optimized multimedia delivery. In the quest for flexible, cost- and quality-optimized mobile multimedia service delivery, cloud computing, Network Functions Virtualization (NFV), and Software Defined Networking (SDN) technologies, some of the key components of the 5G architecture, appear promising. These technologies make it possible to offer on-demand, scalable and elastic content delivery services with performance guarantees, over potentially fragmented physical infrastructures provided by a federated networked cloud [2].

Ideally, this federated infrastructure will involve cloud resources at the mobile edge, thus bringing services as close to end users as possible. For example, video caches running as virtual servers at the edge cloud can improve user experience by minimizing startup latency, while at the same time saving on core network resources, since content is served to users by a local service instance. The ETSI standard on Multi-access Edge Computing (MEC) may play an important role in this direction [3], [4]. MEC allows harnessing the power of cloud computing by deploying application services at the edge of the mobile network, e.g., at the Radio Access Network (RAN) level [5]. This can facilitate content dissemination within the access network and can offer new business opportunities by integrating Mobile Network Operators (MNO) into the video delivery value chain.

In this article, we address the challenges of mobile video delivery in a 5G network context. We present an architecture which allows content providers to deploy virtual CDN instances dynamically over a federated multi-domain

- T. Taleb and I. Benkacem are with the Department of Communications and Networking, School of Electrical Engineering, Aalto University, 02150 Espoo, Finland. T. Taleb is also with Oulu University, Oulu, Finland and Sejong University, Seoul 143-747, South Korea. (E-mail: talebtarik@ieee.org)
- P.A. Frangoudis and A. Ksentini are with the Communication Systems Department, EURECOM, 06410 Sophia-Antipolis, France. (E-mail: {pantelis.frangoudis,adlen.ksentini}@eurecom.fr).

edge cloud. Our system exposes a northbound API over which customers can request the creation of *CDN slices* for a specific duration. These CDN slices consist of Virtualized Network Functions (VNF) deployed over the multi-domain cloud, such as virtual streaming servers, caches, and transcoders, appropriately chained and configured with optimally-assigned resources (e.g., CPU and storage) for specific service-level performance targets. Cloud resource allocation decisions and elastic resource management, so that a virtual CDN instance is scaled following end-user demand, are some of the main issues our design tackles. Resource allocation decisions are measurement-driven: with respect to compute resources, we use empirical models of video QoE as a function of service workload, while service demand information (e.g., maximum number of simultaneous video streams in an area covered by a specific macro-cell) submitted by the customer at service instantiation time can be used to appropriately provision the CDN slice with radio access and core network resources.

Multi-level monitoring information, both at the cloud infrastructure and the CDN slice service/application level, is necessary for elastic slice resource management. Our design provides the support and the appropriate building blocks to collect and utilize this information, and it does so in a standards-based way: our architecture aligns with recently proposed standards for edge computing and NFV, and the ETSI MEC [6] and NFV Management and Orchestration (NFV-MANO) [7] frameworks in particular.

This article is structured as follows. § 2 reviews relevant works in the literature. § 3 presents our vision and architecture for the provision of virtual CDN slices over multi-domain cloud infrastructures in a 5G network context. We then propose schemes for QoE-driven cloud resource allocation (§ 4) and elastic resource management (§ 5). In § 6, we devise and evaluate a model and algorithm for the Elasticity Decision Maker component of the proposed elastic resource management framework. The process of initiating the transcoding service on the fly, as part of the envisioned elastic resource management framework, is described in § 7. Testbed experiments on the performance of the overall CDNaaS system follow in § 8. We conclude the article in § 9.

## 2 RELATED WORK
### 2.1 Network Functions Virtualization

Network Functions Virtualization (NFV) is becoming a key technology for future large-scale service delivery [8]. NFV involves carrying out in software networking tasks that were traditionally performed by costly, special-purpose hardware. Significant standardization efforts are currently taking place around NFV. The European Telecommunications Standards Institute (ETSI) has specified a Management and Orchestration framework for NFV (NFV-MANO [7]), and one of the proposed NFV use cases is the provision of virtualized CDN services [9, Use Case #8].

As we show in § 3.3, our work is aligned with this framework and this use case in particular, with all the basic components of a CDN slice in our design (i.e., caches, load balancers, streaming servers, transcoders, and name servers) being implemented as VNFs. However, NFV is being applied to a diverse set of functions [9]. In this spirit, an effort worth noticing is T-NOVA [10], [11], an EU-funded project proposing a VNF marketplace, where VNF providers will be making available their functions to be deployed over the infrastructure of a network or cloud service provider, developing the necessary support for VNF brokering, management and service delivery.

### 2.2 Multi-access Edge Computing for video delivery

MEC has been proposed as an enabler for novel, low-latency services in a mobile network [5]. Considering its potential, both industry and the research community are working on maximizing the benefits and efficiency of MEC technology. As discussed in [3], [4], [12], such a resilient decentralized architecture will enable new services and promising business models, including those for a smart city [13].

MEC and NFV are complementary technologies, sharing common concepts such as the existence of a virtualization infrastructure where applications can be launched and which is managed by a Virtualised Infrastructure Manager component. As such, the MEC and NFV platforms could be running independently or could share some reusable components (e.g., parts of the virtualization infrastructure and its management utilities). A standardized MEC framework is the subject of a recent ETSI specification [6]. The commonalities in the characteristics of NFV-MANO and the MEC management and orchestration system and the lack of a unified way to jointly manage VNFs and MEC applications have been identified by Sciancalpore et al. [14] who propose specific extensions to the MANO design to address this issue. A basic premise of our work is also the ability to jointly manage resources and virtual instances both on top of and outside the mobile edge, and our design supports this in a standards-compliant way. However, we also need to account for resources leased on heterogeneous cloud infrastructures. Therefore, our architecture includes an extra tier outside the MANO framework, where our CDNaaS Service Orchestrator operates, using standard interfaces to communicate with NFV and Mobile Edge Orchestrators.

Beyond architecture design issues, and focusing on the specific video streaming service, Fesehaye et al. [15] propose a two-hop edge scheme which reflects the data transfer rate and throughput of the edge in comparison with the remote cloud. Fajardo et al. [16] introduce a network-assisted adaptive streaming application to enhance the QoE of the delivered multimedia content. An architecture with distributed parallel edges to increase QoE for content delivery has been proposed by Zhu et al. [17]. Chang et al. [12] deploy independent small-scale data-centers at the network edges, which are capable of performing video caching and streaming on their own. Jararweh et al. [18] integrate caching with proxy functionality at the edge to store media content. They also enforce computation offloading to increase the lifetime of mobile devices.

Video transcoding in the cloud has recently received significant research attention [19], [20], [21]. Utilizing virtual instances on the cloud to perform video transcoding upon request has been proposed in [22], [23] as the simplest and most straightforward use case. The works in [24] and [25] also propose cloud-assisted video transcoding. Utilization of cloud resources to assist mobile devices for customized transcoding services [26] and for energy conservation on mobile devices [27] has also been proposed. As an efficient way of video transcoding in the cloud, an approach to reduce the bit-rate of the transcoded video by using a higher quantization parameter without reducing the frame size or the frame rate has been proposed by Johkio et al. [28]. Transcoding only a portion of a video to reduce the transcoding time [29], [30] and distributed video transcoding in the cloud to enhance efficiency have also been studied in [31], [32]. Amazon has also recently introduced an elastic transcoder [33].

In all the above research works, MEC is considered a promising solution for handling video services, although mainly focusing on streaming, caching and compression techniques. The computationally-intensive transcoding functionality has been generally proposed to be treated on traditional clouds, with the exception of Beck et al. [34], who study video transcoding at the edge for a Voice over Long Term Evolution (VoLTE) service.

### 2.3 5G slicing

There is wide consensus that the 5G mobile network technologies will be more than new radio access schemes [35], [36], with network programmability and flexible service composition being some of their key characteristics. 5G will offer the opportunity to dynamically build and manage end-to-end network slices [37]. A slice can be considered as an autonomous network service instance, encompassing all service-specific functionality as well as the respective resource management [38]. Slices coexist over the same physical infrastructure by means of Cloud, NFV and SDN technologies. Resource management per slice, but also per participating network entity is a critical aspect. Li et al. [39] therefore differentiate between vertical and horizontal slices and the resource management aspects therein. The vertical case refers to resource sharing among different end-to-end services and applications, while the horizontal case focuses on resource sharing among network nodes and devices typically at the edge (e.g., nearby devices sharing computation capacity).

Network slices are built by combining various reusable network functions implemented as software components. In this direction, Nikaein et al. [40] propose a service-oriented network architecture, introducing the term of a *network store*, which serves Mobile Network Operators, enterprises, and over-the-top third parties as a repository of such network functions and network slice templates.

Various types of value added services can be delivered over network slices. We focus on content delivery and, to the best of our knowledge, this work is the first to deal in depth with this application case in this 5G slicing context.

### 2.4 Our prior work

In our prior work, we presented architectures for the provision of generic services on top of a telco cloud. One such service could involve CDN functionality on demand. In particular, we presented the design of an Anything-as-a-Service scheme for 5G networks [41], as well as an architecture for on-demand virtual CDN deployment over a telco cloud [42]. However, in the former case, we did not expand on the very technologies, algorithms and mechanisms for CDNaaS provision, while the latter design was tailored to a single telecom operator and did not consider the particularities of a mobile network; as such, it is not directly applicable to multi-domain 5G settings where services are deployed on the mobile edge cloud and/or to multi-cloud cases. We should note that elastic compute resource management (§ 5) and dynamic video transcoding functionality (§ 7) have been treated independently in our prior work [43], [19]. In this article, we address them jointly, integrating them in our CDNaaS architecture. We improve their design, and present new models and algorithms, fine-tuning their operation. Importantly, the work presented in § 4-6 builds on a measurement methodology and experimental results on video QoE and its dependence on service workload that we presented in [42]. Here, we derive empirical models of QoE as a function of workload out of these experimental data, and use them as a basis for our algorithms for compute resource allocation and elastic resource management, thus demonstrating a practical use case for them. Finally, in another line of relevant work, we focused on VNF placement algorithms for CDNaaS [2], a subject outside the scope of this article.

## 3 A 5G-ORIENTED ARCHITECTURE TO SUPPORT MULTI-DOMAIN CDN SLICES

### 3.1 Features

We propose an architecture tailored to 5G network settings, which allows the dynamic creation and management of CDN slices, automatically reserving the necessary computing, networking, storage and other resources across a federated cloud infrastructure, and dynamically scaling them following changes in user demand and network conditions, as the latter are detected via service and infrastructure level monitoring. The distinctive characteristics of our approach are the following:

- Mobile CDN slice resources can be allocated over infrastructures across multiple domains, thus potentially managed by different parties. These infrastructures include 5G radio access nodes, edge clouds of mobile network operators, and global-scale Infrastructure-as-a-Service (IaaS) providers (e.g., Amazon, Azure, and RackSpace).
- Resource management follows a measurement-driven approach: At the compute level, we use empirical models of the relationship between video QoE and service workload for CPU resource allocation and scaling. At the RAN level, per user video QoE estimates, potentially coupled with (per user and aggregate) network state information exposed by MEC-level services, are utilized for on-the-fly content transcoding decisions to match current radio conditions and optimize user experience.
- Our architecture is aligned with the recent ETSI NFV MANO [7] and MEC [6] standards, and the design choices we have opted for are particularly targeted to multi-domain settings.

Such a service can be offered by a MNO, combining its own centralized and edge cloud with other federated cloud resources. However, with the availability of MEC-level APIs and the ability to deploy arbitrary virtual services on the edge cloud according to specific Service-Level Agreements (SLAs) with the network operator, third parties could offer the CDNaaS functionality and manage mobile CDN slices.

## 3.2 Design components

### 3.2.1 CDNaaS Orchestrator (CDN-O)

The CDNaaS Orchestrator is responsible for generating an abstract representation of a CDN slice instance. Each slice is composed of a set of primitive VNF instances (e.g., streaming servers, caches, and transcoders) appropriately chained together to form a service instance. These instances, their actual placement in specific points in the underlying cloud infrastructures (e.g., edge cloud compute nodes) [2], connectivity properties, as well as the necessary resources to be allocated to them are encoded in an abstract way in the form of a Service Instance Graph (SIG). The SIG is transformed to an actual slice deployment by interfacing with cloud controllers managing the underlying federated cloud.

Much of the resource allocation intelligence is provided by the CDNaaS orchestrator, since this is the component that will have to decide the amount of resources to be reserved per VNF instance (i.e., virtual CPUs, storage, memory), optimize VNF placement and allocate adequate network capacity vertically across the networking technologies that are involved (i.e., RAN and core network level).

The CDN-O exposes a northbound REST API to offer an entry point to customers (content providers) to request the setup of a CDN slice with specific performance requirements and dimensioning information (e.g., maximum expected number of streams per region), as well as handling pricing and SLA-related procedures.

From an implementation perspective, an aspect that increases the complexity of the CDN-O is the requirement to be able to communicate with heterogeneous clouds. At its current version, our software prototype supports OpenStack, AWS, and Microsoft Azure.

Performance-wise, in our implementation, the initial calculation of the amount of CPU resources to assign to cover a specific demand can be executed very fast, in 3 simple steps, as described in § 4.3. How to distribute these resources to VNFs and how to place the latter on edge or other hosts, a topic that has received significant research attention, is not within the scope of this article. In our prototype, we have experimented with simple and very fast to execute placement schemes, but also with more sophisticated algorithms, such as the ones proposed in our prior work [2]. These algorithms have been shown to run in the order of a few seconds or less, for large problem instances.

Eventually, the time to serve a virtual CDN instantiation request critically depends on the size of the deployment in terms of VNF instances. As we have experimentally shown [42] (albeit on a different environment, service architecture, and implementation), this time is dominated by the time it takes to launch the VNF instances (which includes also the time to copy the VM images from the Virtualized Infrastructure Manager (VIM)'s image store to the selected compute nodes); the delay due to the execution of API calls over the customer-facing northbound API or to communicate with the underlying VIMs on the southbound is negligible.

### 3.2.2 CDNaaS Slice Coordinator (CDN-SC)

Each slice needs to be autonomously monitored and managed, particularly given their heterogeneous performance requirements. CDN-service-level functionality is implemented here. For example, the number of video streaming sessions per slice is monitored by specific tools available by the CDN-SC and decisions to scale up to more CPU resources to sustain the desired video QoE according to the slice SLA are taken by this component. The CDN-SC may need multi-level information to optimize the operation of a CDN slice. For example, it may need RAN-level information from a specific macrocell or VNF CPU load information from the edge cloud, which can be retrieved by using specific MEC-level APIs via communicating with specific edge management components.

### 3.2.3 Mobile Edge Orchestrator (MEO)

We are targeting multi-domain scenarios, where different MNOs contribute their resources at the mobile edge in a federated cloud. Each MNO therefore operates its own Mobile Edge Orchestrator (MEO) component, and the CDN-O and CDN-SC need to interface with multiple MEOs for slice deployment and management. Each edge node may be hosting VNF instances belonging to different slices. It is the role of the MEO to maintain an overall view of the edge hosts that it manages and the underlying cloud resources (via communicating with the respective edge cloud VIMs), and expose specific APIs to the CDN-SC and the CDN-O to instantiate slice components (as virtual machines on top of edge hosts) [5]. The ETSI MEC framework and reference architecture [6] identify specific management reference points where these interfaces are implemented. In § 3.3, we provide further details on how our design complies with this standard and how our components interact with the mobile edge system and platform.

## 3.3 Standards compliance and extensions

### 3.3.1 Mapping to the ETSI NFV MANO and MEC reference architectures

Fig. 1 presents a simplified view of the MEC and MANO frameworks and shows how our main architectural components map to them. Since our architecture is designed to operate in a cross-domain fashion, the CDN-O rests outside the MANO and MEC domains, and, depending on the deployment scenario, can be considered either as part of the network operator's Operations Support System (OSS) or as a separate third-party entity. It communicates with the NFV Orchestrator (NFVO) and the MEO over the Os-Ma-nfvo and the Mm1 interfaces, respectively.

Service-level components of a slice (e.g., cache or transcoder virtual instances) can be instantiated as VNFs, Mobile Edge (ME) applications (equivalent to VNFs, in the MEC terminology), and/or virtual instances on top of third-party, non-MANO-compatible infrastructures. In the latter case, the CDN-O uses the cloud provider's APIs to deploy them (e.g., OpenStack or Amazon EC APIs).

The CDN-SC functionality is implemented in the VNF Manager (VNFM) component of the MANO framework. To manage a CDN slice's life-cycle, it interacts with cloud VIMs, but also monitors and manages slice components running as ME applications. Specific services running at the Mobile Edge Platform (MEP), e.g., the Radio Network Information

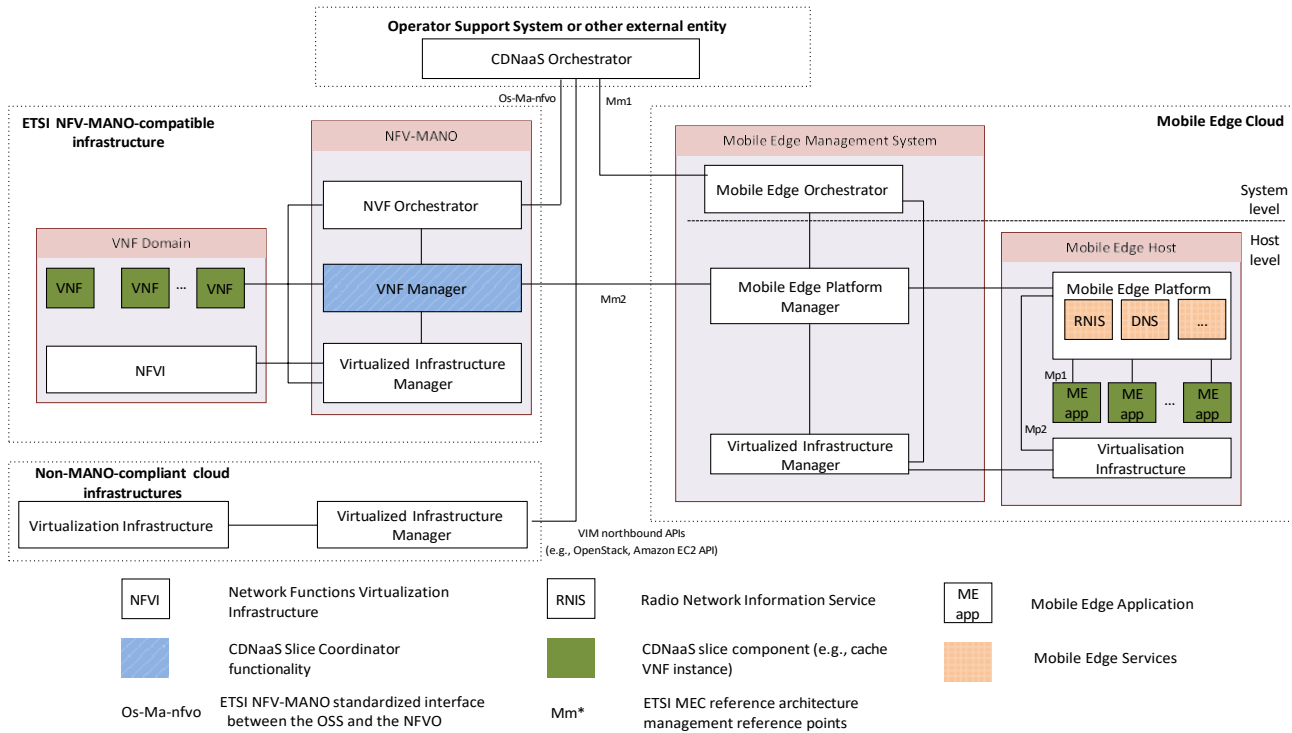

Fig. 1. Our design in the context of the ETSI NFV MANO and MEC reference architectures. The figure includes some of their building blocks and standardized interfaces between them, and the components of our architecture.

Service (RNIS) [44] (see § 5.1), can provide input for runtime adaptations. Note that the use of Mm2 to access the MEP Manager (MEPM) from outside the MEC domain is not strictly complying with the MEC standard. However, since MEPM–CDN-SC communication is necessary in our case, we propose to repurpose this interface by opening it to the VNFM in the same way that it is accessible to the OSS/BSS.

### 3.3.2 CDN deployment for traffic delivery from the edge

CDN traffic in our case is delivered using DNS redirection techniques. When there is no MEC coverage, a user is served by the cache to which its request for content is resolved, be it at a centralized cloud infrastructure (e.g., Amazon) or at a nearby NFVI, e.g., at the closest operator's Point of Presence. When a user is attached to the mobile network and there is MEC support at its location, the local MEP DNS server is used to respond to the user's queries, and traffic may be terminated at a local ME application. In our implementation, MEC clouds are organized in *regions*. Each region corresponds to a specific set of MEC hosts, and one MEO may manage multiple regions. The required steps for specific mobile user traffic to be delivered via the appropriate ME application follow:

**Step 1: Application package preparation.** The CDN-O, which receives the CDN deployment request, prepares for onboarding the CDN components (e.g., video cache) to be deployed at the edge. It creates an application package, which includes an Application Descriptor (AppD) [45] with the resource requirements of the application instances, a pointer to the VM/container image data, and, importantly, traffic and DNS rules (`appTrafficRule` and `appDNSRule` fields of the AppD), so that DNS requests for a specific name are resolved to the IP address of the edge cache application instance, and CDN traffic is served from it.

**Step 2: Onboarding.** The CDN-O, depending on where the customer wishes to deploy the service, accesses a MEO's package onboarding endpoint (HTTP request with the AppD and other information as its body, as specified in ETSI MEC 010-2 [45]). The region is designated in a `RegionID` field which we have added to the original standard specification of the Application Package Onboarding message; based on that, the MEO then selects the responsible local edge VIMs to onboard the package.

**Step 3: Instantiation.** The CDN-O creates the application instance via the MEO's instantiation API endpoint. The region identifier is included in the request, so that the MEO decides which edge VIM should host the instance and locates the API endpoint (Mm5 reference point) of the responsible MEPM. After successful creation of the instance, the MEO notifies the MEPM of the DNS and traffic rules to be added; the MEPM then reconfigures the MEP's DNS service to resolve user queries for specific domain names indicated in the AppD to the instance's IP address, and, optionally, interacts with the data plane to apply further traffic steering rules (in our case, data plane traffic at the edge is handled using SDN, by interfacing with an OpenFlow-capable virtual switch).

### 3.3.3 Support for MEC-in-NFV deployments

Our design treats MANO and MEC as two distinct domains. It is the responsibility of the CDN-O and CDN-SC to handle the deployment and management of CDNaaS service components seamlessly across edge and centralized cloud infrastructures. Our design does not preclude operation in a MEC-in-NFV environment, where the MEP/MEPM are deployed as virtual instances and their instantiation and lifecycle management, as well as those of ME applications, are delegated to an NFVO. The interplay between MEC and NFV is currently under study by ETSI MEC 017 [46], and in parallel research efforts [14].

VNFs and ME applications have some distinct differences, which are reflected in the structure of their descriptors: An AppD, beyond the information included in a VNF descriptor (VNFD), also provides information about MEC services consumed and exposed by the ME application, and specific DNS rules and traffic steering directives. In a MEC-in-NFV scenario, to deploy a CDN component at the edge, the CDN-O would request the MEO (or, MEAO, standing for ME Application Orchestrator, in ETSI MEC-in-NFV parlance) to instantiate the ME application, and the ME(A)O would create a network service descriptor (NSD) for the application to pass on to the NFVO, which would eventually create the instances. However, the definition of the NSD according to ETSI NFV-MAN 001 [7] does not include AppD references. On the other hand, a VNFD on its own cannot capture the requirements of ME applications and the NFVO does not know how to handle them. This is an issue already acknowledged by ETSI.

In our approach, it is transparent to the CDN-O if MEC is implemented in the standalone or the in-NFV version. The integration of MEC in NFV is a matter of a set of implementation choices, to which the CDN-O is agnostic; the latter need only prepare the application component for deployment either at the edge (application package onboarded to the ME(A)O over Mm1) or at generic NFVIs (NSD including VNFD references onboarded to the NFVO via Os-Ma-nfvo). In our solution, the NFVO is left unmodified, while the ME(A)O, which receives an application package from the CDN-O, is responsible for extracting the MEC-specific elements from the AppD, translating it to a standards-compliant NSD with references to VNFDs for NFVO onboarding and instantiation, and handling all the MEC-specific functionality itself. A critical aspect is the interface with the data plane for traffic steering. We opted for a solution where the Mp2 reference point is maintained as a MEC-internal, proprietary interface, to which NFV MANO is agnostic. (The other option would require the ME(A)O to translate traffic rules to a Network Forwarding Path (NFP), and pass it to the NFVO to apply it at the VIM. We did not select this option due to its additional complexity for the ME(A)O.) Finally, regarding the communication of the CDN-SC with the MEPM (e.g., to update traffic rules) and the ME applications running as VNFs, the new Mv2 and Mv3 reference points can be used, respectively.

### 3.3.4 MEC support for network slicing

Full MEC support for network slicing is not yet there. The ETSI MEC 024 work item [47] has recently been established to study this issue, providing some early use cases and requirements [48]. Especially in a MEC-in-NFV environment, multiple virtual MEP/MEPM instances might coexist, each handling a different slice or set of slices. An issue that then emerges is how to discover the MEP instance responsible for an application at instantiation time.

We have implemented extensions to offer basic MEC slicing support. To manage the logical grouping of ME applications and other instances that are part of a slice, and to give hints to the ME(A)O on how to handle special types of instances (e.g., virtualized MEP/MEPM), we exploit placeholder fields in the information model specified in ETSI MEC 010-2 for the MEO-OSS interface; this way, we limit the impact of our extensions on the standard. At package onboarding, we use the `userDefinedData` field of the Application Package Onboarding Request to signal the MEO that the package contains a virtualized MEP. Then, at application instantiation, we use the `selectedMEHostInfo` element to add slice identification information and to define the regional data center where the instance is to be deployed. The MEO then uses the slice ID to select the appropriate virtualized MEP/MEPM instance, and communicates with it to discover the API endpoints of its services, register the application, configure traffic steering rules, etc.

## 4 MEASUREMENT-DRIVEN COMPUTE RESOURCE ALLOCATION

A critical aspect in the deployment of a CDN slice is to provision it with adequate compute resources to sustain the customer's desired end-user QoE levels. This decision has direct implications for the CDNaaS operator regarding the cost of the slice deployment, since this is a function of the cloud resources leased. It is also directly related with the SLA between the customer (content provider) and the system operator and with service pricing: The price of a service instance is a function of the amount of resources used, and the latter (and their elastic management) are important to provide the customer with service-level guarantees.

In order to decide on the optimal amount of compute resources to allocate to a slice given customer-defined demand and QoE specifications, we follow a measurement driven approach. In our prior work [42], considering specific technologies at the VIM and the CDN service level, we carried out testbed experiments where we measured how QoE for a virtualized video service is affected by service workload. In this article, based on these experimental results, we derive an empirical model of QoE as a function of this workload, which we apply to (i) decide on an initial resource dimensioning, i.e., how many compute resources to dedicate to VNF instances composing a CDN slice, (ii) dynamically scale these resources in real time, so that the target QoE levels are attained, and (iii) be able to identify, based on data from real-time slice monitoring, what is the root cause in case of service quality degradation and act appropriately (i.e., identify whether QoE degradation is due to CPU load or reduced network capacity). We provide a brief overview of our experimental methodology, before we present how the results of our tests are utilized in this work.

### 4.1 Measurement methodology

We focus on a video streaming service where video content is delivered using Dynamic Adaptive Streaming over HTTP (DASH) [49]. The use of HTTP is typical for delivering mobile video over CDN infrastructures nowadays. Under DASH technologies, the client receives a Media Presentation

Description (MPD) file with information on the available representations (different qualities) of the same video, which is segmented in chunks. Afterwards, the client proceeds to download the video chunk-by-chunk, potentially switching among the available qualities (and thus bit-rates) to better adapt to the network conditions.

To quantify the relationship between server load and user experience, we measure the performance capabilities of a standard NGINX HTTP server [50] hosted on a KVM [51] virtual machine, pinned to a single CPU, and in the presence of multiple parallel video sessions. We selected KVM since it is the default hypervisor in OpenStack.[1] We use the terms virtual CPU (vCPU) and CPU core interchangeably; in a cloud setup, this corresponds to a 1:1 CPU allocation ratio.

Our QoE metric is the Mean Opinion Score (MOS), i.e., the expected rating that a panel of users would give to the quality of the transmitted video in the 1-5 (poor-excellent) scale. To estimate it, we apply the Pseudo-Subjective Quality Assessment (PSQA) approach [53]. PSQA consists in training a Random Neural Network (RNN) based on experiments with physical subjects under controlled conditions, where a set of parameters affecting quality is monitored and the ratings of users are recorded. The trained RNN classifier can then be applied in real time and output the expected MOS for specific values of the input parameters. Singh et al. [54] have applied PSQA to estimate QoE for H.264/AVC-encoded video delivered over HTTP, and we have used their tool in our experiments. PSQA receives as input playout interruption statistics and the average value of the Quantization Parameter (QP) across all picture macroblocks in a QoE measurement window, which corresponds to 16 s of video. QP is an indication of picture quality (the higher the QP, the lower the video bit-rate and quality). More details on our experimental methodology can be found in [42].

### 4.2 Empirical model of QoE as a function of service workload

With the aim of empirically associating video server load (in terms of parallel video streams) and user experience, we varied the number of parallel video sessions from 2000 to 12000. Fig. 2 presents the average MOS value observed across all 16 s video samples for each case. We observe that a vanilla NGINX server can sustain up to more than 5000 parallel HD video sessions with an excellent quality. For loads of 6000 parallel users or more, video interruptions start to take place and this reduces QoE, especially as load grows. The duration and the frequency of such playout interruptions follow an increasing trend with server workload, reaching, on average, more than 7 s and 2.5 per minute respectively when the number of parallel streams is 12000 [42].

We then fit a quadratic function to our measurement results. We found that the following expression reasonably describes our data (the coefficient of determination is $R^2 = 0.9265$) and approximates the average QoE as a function

---

1. In [42], we also ran experiments to compare the performance of KVM and containerization technologies (Docker [52]) for video and generic HTTP services and found the latter to achieve superior performance. Such a comparison is outside the scope of this work; our methodology and algorithms are directly applicable to Docker as well.

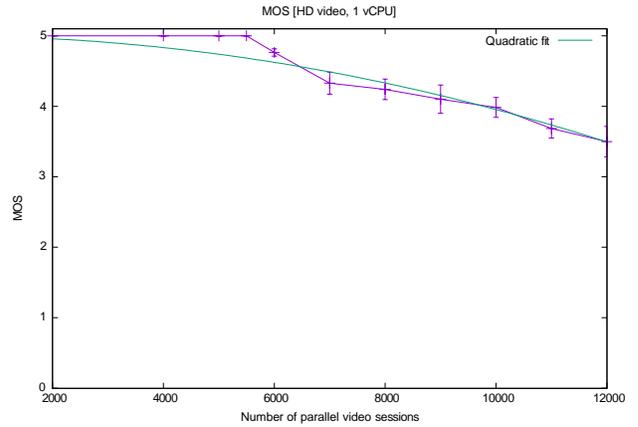

Fig. 2. Average QoE as a function of the number of parallel HD video streams on a web server hosted in a KVM VM using 1 CPU core. Each point is the mean of a few hundreds of QoE samples (MOS values), presented with 95% confidence intervals. A quadratic function has been fit to the empirical data.

of the number of parallel video streams on a single-vCPU HTTP server virtual instance:

$$Q(x) = 5 - 1.046 \times 10^{-8} x^2, \qquad (1)$$

where $Q(x)$ is the expected QoE of a user expressed in MOS terms when there are $x$ parallel video streams delivered by the same virtualized server.

Our experiments also revealed the linear performance scalability as the number of vCPUs allocated to a virtualized HTTP server instance grows: We found that by adding one more vCPU to the instance, it could handle double the number of parallel video streams with no interruptions.

We also report on the CPU utilization ratio as service workload grows. Fig. 3 shows that for more than 6000 parallel streams (1 vCPU case), CPU utilization in the virtual machine that hosts the video server approaches 100%. This is the point when playout interruptions start to take place. It should be noted that with 2 vCPUs, we did not manage to saturate the VM, since we were hitting network-level bottlenecks before processor-level ones. This result suggests that CPU load can be used as a further indication of QoE and can be used as an additional metric when taking CDN slice resource scaling decisions. We propose such scaling mechanisms in Sections 5 and 6.

### 4.3 CDN slice compute resource calculation

Using the northbound REST API of the CDN-O, the customer provides the desirable maximum demand in terms of the number of parallel video streams and the target QoE level, among other information. During the operation of the slice, end users will be served content by the optimal video server, which is reasonable to assume that it will be their local cache.[2] The number of vCPUs to be allocated to the cache VNF instances of a slice is then calculated in a straightforward way using the following procedure:

---

2. In our implementation of a CDNaaS scheme, requests for video content are served by caches acting as user proxies towards origin video servers, from which they retrieve content if they do not have it stored locally.

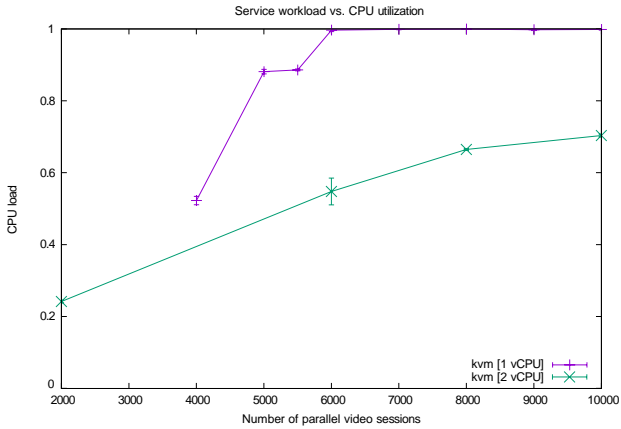

Fig. 3. CPU utilization as a function of video service workload (parallel number of streams). Each point is the mean of a few hundreds of QoE samples (CPU load values), presented with 95% confidence intervals.

**Step 1:** Get customer demand in terms of the maximum number of streams ($n$), and the minimum tolerable QoE threshold ($q_{min}$).

**Step 2:** Find the maximum number of streams a vCPU can handle with a QoE value greater than or equal to $q_{min}$. To derive this value, we use the inverse function of Eq. (1), which gives us a number of parallel streams ($n^*$) for the given $q_{min}$ value. In our empirical model, calculating the inverse of $Q(x)$ is straightforward.

**Step 3:** The number of vCPUs to allocate for a slice is $v = n/n^*$. Namely, we divide the total demand by the maximum number of streams a vCPU can sustain with a given QoE.

Note that $v$ refers to the sum of the compute resources of the slice. These have to be allocated to VNF instances and the slice has to make sure using the appropriate CDN components (i.e., load balancers and redirection servers) that simultaneous requests for video content at a VNF instance with a specific number of vCPUs will not exceed its processing capacity, to guarantee the desired QoE threshold. It is reasonable to assume that the customer will be able to submit a CDN deployment request with a different target demand per region (as defined, e.g., by the coverage area of a macrocell or a regional data center; the granularity of a region is up to the service operator to define). Then, the above procedure can be executed once per such region.

## 5 SLICE MONITORING AND ELASTIC RESOURCE MANAGEMENT

Real-time slice resource management decisions require multi-level service monitoring. In this section, we detail the monitoring-related components of our architecture, and how they can be used to dynamically and elastically adjust the allocated resources to meet the desired performance criteria.

We identify two major types of resources that affect user experience in a video content delivery setting: cloud and network related ones. In the first case, we focus on compute resources, while in the second case, we are mostly interested in the conditions in the mobile edge (RAN level) and attempt to adjust the delivered video bitrate on a per user basis to address network capacity and link quality changes. We introduce the following components at the CDN-SC level: Resource Usage Monitor, Quality Assessor and Elasticity Decision Maker.

### 5.1 Resource Usage Monitor (RUM)

This component is responsible for monitoring resource usage for all VNF instances of a slice. Such information can be provided by the VIM. For example, if OpenStack is used as the VIM platform, as in our implementation, resource usage information per VM can be provided by OpenStack's Ceilometer module. RUM periodically monitors every instance's CPU and RAM utilization (other parameters are monitored too, but are not considered in the methods we are describing in this work) and communicates this information to the Elasticity Decision Maker.

The RUM is also responsible for collecting network-level information. For mobile users and in regions where CDN components are deployed as ME applications, the RUM has access to information on the radio network conditions via the RNIS [44]. Our implementation includes a standards-compliant RNIS (see [55] for details on its internals), which allows ME applications to subscribe to radio information at different granularity (i.e., per user or per cell), and receive them as asynchronous notifications. The RUM can thus, at any time, be informed of the load of a cell in terms of connected users, but also of the Channel Quality Indication (CQI) of each user consuming the CDN service. This information can be used in two ways. First, significant changes in the CQI or cell load values can be quickly detected and can trigger the initiation of video transcoding. Second, the CQI values, combined with the cell load, can be used to estimate the radio capacity per user. This, in turn, can drive the video transcoding process. This estimation can be achieved by mapping CQI values to achievable throughput for different cell loads via experiments, or using information provided in the standards. The 3GPP TS 36.213 specification [56] provides a mapping of CQI values to modulation and coding schemes (MCS) that a base station uses for transmitting data to users, as well as a translation of the MCS to achievable throughput for a specific amount of radio resources allocated to a user. Combining this value with the number of coexisting users can provide a rough estimate of a user's radio capacity.

Note that from an architectural viewpoint, applications running outside the edge computing environment cannot directly subscribe to MEC services, nor have direct access to the MEPM. A potential solution is to expose the MEPM Mm2 interface to specific external entities such as the RUM, which is part of the CDN-SC (VNFM-level functionality). Otherwise, the RUM can operate a monitoring component as a ME application authorized to consume the RNIS and relay the collected radio information to the RUM. This approach extends to other types of monitoring information which are not directly accessible by external entities via MEC APIs. Although the first approach necessitates to repurpose the Mm2 interface, it is simpler to implement.

### 5.2 Quality Assessor (QA)

System level information is important for resource management decisions, but it does not offer a feel about how users perceive the CDN service. We are interested in having a user-centric view about the experience of the end-users of the CDN slice and, in turn, perform QoE-aware management decisions. To this end, we have introduced the Quality Assessor component, which can estimate the CDN slice performance in QoE terms. Our proposed scheme thus decides on whether to scale resources or not also based on the QoE perceived by users, as estimated and reported by the QA.

The QA has two parallel operation modes. First, it operates as a QoE probe and its purpose is to get an overall view of the QoE that can be delivered by a VNF instance from a compute perspective. Namely, the probe's purpose is to detect any degradation related with heavy workloads and not individual user characteristics (e.g., link quality) or network capacity issues. As such, it can be launched in the same host or data center as the monitored VNF instance(s) (to minimize network-related effects) and emulate a user which receives a video stream from the specific monitored VNF instance. The probe records QoE-relevant parameters, uses the PSQA model to calculate a QoE estimate (MOS value), and reports it to the Elasticity Decision Maker. A simpler, less accurate implementation option is to monitor the number of parallel video streams served by the instance and use the model of QoE vs. workload derived in § 4 to estimate a MOS value.

The second mode has to do with individual user QoE monitoring. It works under the assumption that at the user end there is the appropriate functionality in place to record the relevant input parameters for QoE calculations and report them to the QA.[3] The QA receives real-time information of the service status from the client. The status includes the ID of the last downloaded segment and of the playing segment, the playback interruption count and duration, and the QP value of the video playing at the end-user device. Considering these parameters, the PSQA model is used again to generate a MOS estimate in real-time in an automatic manner for a specific user. If the MOS value is below the target threshold set by the customer, the QA needs to verify that this is an individual user issue or that this is due to heavy load. This information is available by the RUM. In case resource utilization is within the acceptable thresholds, the QA triggers the CDN-O to initiate the video transcoding service. The latter then communicates with the MEO of the edge network serving the specific user to instantiate the appropriate transcoder VNF on the edge cloud. The transcoding process is presented in detail in § 7. In a different case, this is a problem that affects all users served by a VNF instance and will be tackled by the Elasticity Decision Maker by scaling up the CDN slice.

### 5.3 Elasticity Decision Maker (EDM)

The main EDM functionality consists in deciding when to trigger the CDN-O to enforce the elasticity operation on an instance, indicating what resources should be allocated for the instance while preventing service interruption. The EDM periodically receives monitoring information from both the RUM and QA. Based on that, it identifies the appropriate amount of resources to (re)allocate to an instance to meet the required service level objectives in terms of QoE, also avoiding cloud resource underutilization. The EDM employs a Multi Attribute Decision Making (MADM) algorithm using multiple inputs, such as the resources allotted for an active instance, those in use, maximum resources that can be allocated for the service, instance flavor size,[4] the available flavor types/sizes, and the MOS of the running VNF instance as reported by the QA. The thresholds of each of these inputs are preset for variable flavors and are tuned offline taking into account the results of our experiments (see § 4) ensuring optimal operating conditions.

When EDM takes a resource re-allocation decision, it triggers up-/down-scaling of a running instance specifying its new size/flavor. Scaling refers to increasing or decreasing the resources in terms of CPU, RAM, and storage. It can be either vertical or horizontal. Through vertical scaling, the CDN-O increases the computing power on a running instance, while horizontal scaling adds computing power by adding more virtual instances. Whilst horizontal scaling is relatively straightforward, it is costly as it involves more VMs and is more complex from a management perspective. Furthermore, horizontal scaling requires load balancers to be more effective. In contrast, vertical scaling exhibits less operation cost but is more complex from an implementation perspective. In this work, we report on results obtained in the case of vertical scaling, as its horizontal counterpart is relatively easier to implement.

The EDM compares the CPU usage and RAM usage (RUM output), and the MOS estimate (QA output) against their respective threshold values, and decides on how to elastically adjust the allocated resources as follows:

1) If resource usage exceeds the maximum defined threshold and the MOS score is below the optimal value, the EDM initiates a vertical scale-up.
2) If resource usage is below a minimum threshold while MOS is acceptable, the EDM initiates a vertical scale-down.
3) If resource usage is within the threshold range but the MOS value is below the optimal value, the EDM identifies that the root cause of QoE degradation is not due to increased load; thus, no scaling is decided.
4) If resource usage is within the threshold range but the MOS value exceeds the optimal value, EDM takes no action, deeming the system healthy.

It should be noted that scaling decisions should take into account the capacity constraints of the physical infrastructure. At a low level, such capacity information is available by the VIMs and is passed on to the EDM by the RUM. The interactions between the RUM, QA, EDM, and the rest of our CDNaaS architecture are shown in Fig. 4 following the algorithm proposed in § 6. Fig. 5, on the other hand, focuses on individual user monitoring by the QA, and how the latter interacts with a MEO to initiate the transcoding service (§ 7).

---

3. In practice, much of the video nowadays is delivered via web-based players. Technologies such as the HTML5 media source extensions, which are becoming widely supported, facilitate the development of custom web players for HTTP video (see, e.g., the DASH Industry Forum reference client, a pure Javascript/HTML5 MPEG-DASH client implementation [57]). This may make the requirement for additional client-level monitoring functionality easier to satisfy.

4. We use the term flavor as in OpenStack, to represent a block of fixed resources in terms of CPU, RAM, storage, ephemeral disk and swap memory. When a flavor is allocated to an instance, a certain amount of resources that remain fixed during its operation are assigned to it.

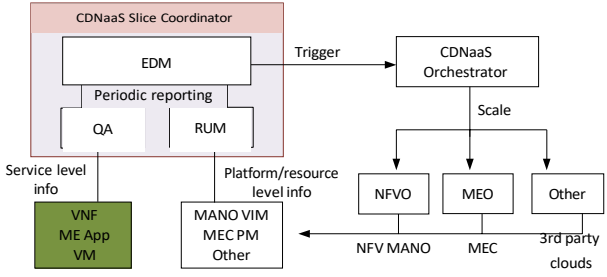

Fig. 4. Cloud resource and QoE monitoring components at the CDN-SC level and their interactions.

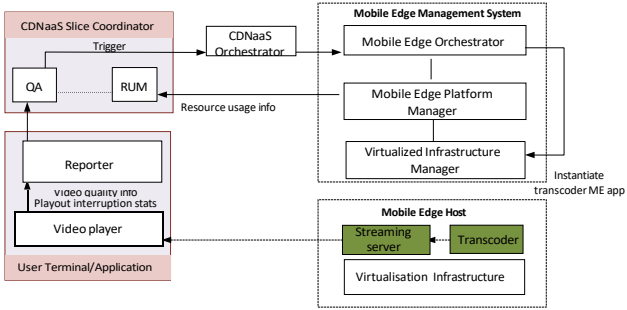

Fig. 5. Individual user QoE monitoring functionality of the QA and transcoding instantiation on top of the edge cloud.

## 6 EDM ALGORITHM

The EDM component monitors the virtual resources of all VNFs composing a CDN slice. In case of under/over-utilization, it triggers the CDN-O to enforce the elasticity operation, specifying the new resources allocated for it (i.e., vertical scaling) or adding new VNF instances (i.e., horizontal scaling) to ensure the desired QoE. The underlying EDM resource algorithm aims to ensure a balanced session load over different instances and to guarantee the highest QoE required by the slice end-users, while minimizing the cost in terms of pricing (e.g., reducing the number of created VNF instances). The EDM periodically receives monitoring information from both (i) the RUM component, that monitors every instance's CPU and RAM utilization and communicates this information to the EDM, and (ii) the QA that estimates the CDN slice performance in terms of QoE. Hence, the EDM algorithm can perform elastic management decisions based on load and QoE awareness.

### 6.1 Mathematical modeling of the EDM algorithm

The proposed EDM algorithm is fed with the aforementioned information in order to identify the appropriate amount of resources to (re)allocate to an instance to meet the required service level objectives in terms of QoE and cost, and to also avoid cloud resource underutilization/overutilization. In this section, we model the physical network representing the cloud infrastructure of a CDN slice. In this regard, we define $\vartheta = \{v_1, v_2 \ldots v_n\}$ as the set of running VNF instances hosted on different public and private IaaS clouds. We define $\xi = \{\zeta_1, \zeta_2 \ldots \zeta_K\}$ as the set of all available cloud domains. We denote by $F = \{f_1, f_2 \ldots f_p\}$ the set of available flavors across all administrative cloud domains, whereby $f_{i,cpu}$, $f_{i,ram}$, and $f_{i,cost}$ denote, respectively, the number of CPU cores, memory capacity and the cost (\$/hour) of flavor $f_i$. In other words, each cloud domain $\zeta_i \in \xi$ has a set of flavors $\{f_{\zeta_i,1}, f_{\zeta_i,2} \ldots f_{\zeta_i,k_i}\}$ where $\bigcup_{\zeta_i \in \xi}\{k \in [1, k_i], f_{\zeta_i,k}\} = F$. Since the QA and RUM components update the EDM component on the basis of a time period $T$, each VNF instance $v_i \in \vartheta$ has a set of initial properties. At the beginning of every period, $v_i$ is hosted at a cloud domain $\xi$ with an initial flavor $f^*_i$. RUM also receives the average number of parallel streaming sessions $\rho_i$ and the initial load $l^*_i$ representing the average resource usage in a VNF instance $v_i$ during the time period $T$. QA provides the estimated QoE $Q^*_i$ from the perspective of the end-users of the CDN slice. For an optimal flavor assignment over VNF instances, we solve the problem using a branch and bound method. For each potential assignment $x_{ij}$ of a flavor $f_j$ to a VNF instance $v_i$, we define a set of values in terms of cost, load and QoE. A possible assignment $x_{ij}$ of flavor $f_j$ to VNF instance $v_i$ represents the following information:

- The price $c_{ij}$ in \$/hour of the allocated flavor.
- The estimated CPU load $l_{ij}$ based on the number of active parallel streaming sessions. The load can be considered optimal when $l_{ij} \in [l^{min}, l^{max}]$. In our model, we estimate $l_{ij}$ as ($l_{ij} = \frac{f^*_{j,cpu}}{f_{j,cpu}} \times l_{ij*}$).
- The estimated QoE that can be provided if we allocate

the virtual resources $f_j$ to $v_i$.

TABLE 1
Summary of notation

| Notation | Description |
|---|---|
| $\vartheta$ | Set of running VNF instances. $\vartheta = \{v_1, v_2, v_i \ldots v_n\}$ |
| $F$ | Set of all available flavors across all cloud domains. $F = \{f_1, f_2 \ldots, f_j \ldots f_p\}$ |
| $f_{j,cpu}$, $f_{j,ram}$, $f_{j,cost}$ | Number of CPU cores, memory capacity and the cost (\$/hour) on the cloud provider of flavor $f_j \in F$, respectively. |
| $x_{ij}$ | Binary variable indicating a possible assignment of flavor $f_j$ to a VNF instance $v_i$. |
| $\xi$ | Set of all available cloud domains. |
| $T$ | Time period. |
| $f^*_{ij*}$ | Initial flavor of the VNF instance $v_i$. |
| $l^*_{ij*}$ | Initial percentage value representing the average resource usage in the VNF instance $v_i$ during a period T. |
| $Q^*_{ij*}$ | Initial QoE value collected for the VNF instance $v_i$. |
| $c_{ij}$ | Estimated cost in \$/hour while using a potential flavor $f_j$. |
| $l_{ij}$ | Estimated load of the instance $v_i$ while using a potential flavor $f_j$. |
| $Q_{ij}$ | Estimated quality of experience of the instance $v_i$ while using a potential flavor $f_j$. |
| $Q^{min}$ | Minimum QoE required in a CDN slice. |
| $l^{min}$, $l^{max}$ | Minimum load and maximum load, respectively, for a VNF instance to be considered using optimally its resources. |
| $C^{max}_{\zeta_k}$ | Maximum capacity of a cloud provider $\zeta_k \in \xi$. It is considered as the available virtual resources on the cloud. |
| $\eta$ | Normalization function. |
| $\rho_i$ | Average number of parallel video sessions in the VNF instance $v_i$ during a time period T. |
| $\sigma$ | Impact of the cost of the flavor on QoE. $\sigma$ varies from 0 to 1. |

Under the assumption that performance linearly scales as the number of vCPUs allocated to a streamer instance grows,[5] we can generalize (1) for every possible assignment $x_{ij}$ of $f_j \in F$ to $v_i$ as follows, also attempting to capture the impact of the cost of a flavor on QoE.

$$Q_{ij}(\rho) = 5 - 1.046 \cdot 10^{-8} \times \left(\frac{\rho}{f_{j,cpu}}\right)^2 + 5\sigma \times \eta(f_{j,cost}) \quad (2)$$

In the above expression, we assume that the cost of a flavor has an impact $\sigma$ on QoE (e.g., in case of two flavors with the same virtual resources, the cheaper one will provide a lower QoE). The rationale behind this decision stems from the fact that there are aspects that could be influencing QoE which are not directly captured in our model. For example, more expensive flavors could come with higher-throughput network connections, high-availability guarantees for the respective VMs, etc. Setting this value to zero leads our model to ignore these effects. We normalize the flavor cost function using the normalization function $\eta_k$, where the intention is to allow the comparison of the impact of the flavor cost in different clouds in a way that eliminates proportionally the effect of prices to promote also the expensive IaaS clouds to a notionally common scale altogether. The normalized value varies between 0 and 1.0:

$$\eta_k(f_{j,cost}) = \frac{f_{j,cost} - min(f \in \zeta_k)}{max(f \in \zeta_k) - min(f \in \zeta_k)}$$

In our simulations, we ensure that the number of parallel video sessions does not exceed a maximum value $\rho_{max}$, so that QoE does not take negative values.

The aggregate utility minimization Linear Programming (LP) problem is shown as follows. For the sake of readability, the notations used throughout the paper are summarized in Table 1.

$$\min z = \sum_{i \in \vartheta} \sum_{j \in F} c_{ij} \cdot x_{ij}$$

s. t.

$$\forall i \in \vartheta, \quad \sum_{j \in F} x_{ij} = 1 \quad (C1)$$

$$\sum_{i \in \vartheta} \sum_{j \in F} x_{ij} = n \quad (C2)$$

$$\forall i \in \vartheta, \quad \sum_{j \in F} l_{ij} \cdot x_{ij} \leq l^{max} \quad (C3)$$

$$\forall i \in \vartheta, \quad \sum_{j \in F} l_{ij} \cdot x_{ij} \geq l^{min} \quad (C4)$$

$$\frac{1}{n} \sum_{i \in \vartheta} \sum_{j \in F} Q_{ij} \cdot x_{ij} \geq Q^{min} \quad (C5)$$

$$\forall k \in \xi \sum_{j=k_{init}}^{k_{fin}} f_{j,cpu} \cdot \sum_{i \in \vartheta} x_{ij} \leq C_{\zeta_k}^{max} \quad (C6)$$

$$\forall i \in \vartheta, \quad \forall j \in F, \quad c_{ij}, l_{ij}, Q_{ij} > 0 \quad (C7)$$

$$\forall i \in \vartheta, \quad \forall j \in F, \quad x_{ij} \in \{0, 1\}, \quad (C8)$$

(3)

---

5. We verified this assumption experimentally: By adding one more vCPU to the instance, it could handle double the number of parallel video streams with no interruptions, keeping the same QoE level.

The objective aims at reducing the total incurred cost while ensuring an optimal load and QoE required by end-users receiving video content from all VNF instances. The constraints in our LP model ensure the following conditions:

- (C1) & (C2): ensure that each instance will be assigned one and only one flavor.
- (C3) & (C4): ensure that all selected flavors can handle a load between $l^{min}$ and $l^{max}$, so as to avoid under/over-utilization of cloud resources.
- (C5): ensures that the average QoE in the new resource allocation satisfies the users' requirement. The average QoE of a CDN slice is defined as the mean QoE experienced at the total VNF instances composing the CDN slice.
- (C6): A cloud domain $\zeta_k \in \xi$ has a maximal capacity of resources $C_{\zeta_k}^{max}$, which represents the total number of CPU cores available at its physical hosts. The constraint guarantees that the maximum capacity is not exceeded when allocating the flavors. We must also clarify that flavors are imported in order by cloud where $k_{init}$ is considered as the index of the first flavor of the cloud $\zeta_k$, and $k_{fin}$ represents the index of the last flavor in that cloud. [6]
- (C7): ensures that the incurred cost, QoE and load parameters are valid values.
- (C8): shows binary variables whereby

$$x_{ij} = \begin{cases} 1 & \text{if flavor } f_j \text{ is assigned to VNF } v_i \\ 0 & \text{otherwise} \end{cases}$$

### 6.2 Algorithm operation

The EDM algorithm is executed periodically after receiving updates from the RUM and the QA regarding the resource usage and QoE status, respectively. It derives an optimal assignment $x_{ij}^{opt}$ of flavors to active VNF instances using the model (3). In case an optimal solution is found, EDM will simply make scaling decisions by increasing or reducing the virtual resources in the running VNF instances. However, in case the LP model is infeasible, we aim to figure out which constraints make the solution not possible. We assume that we cannot relax the QoE-related constraint (C5), since maintaining a minimum average QoE is a strict requirement for the satisfaction of end-users. We therefore relax the model only from the load perspective and start by setting the maximal load $l^{max}$ to 100 in constraint C3. If an optimal resource allocation is found, this means the slice will be over-utilizing its resources in order to ensure the minimum QoE required. Thus, EDM decides to add a new VNF instance to the CDN slice and runs again the algorithm with the $(n + 1)$ VNF instances. Then, we also relax the model from the other side and set the minimal load $l^{min}$ to 0 in constraint C4. If a solution is found, this means that the slice will be under-utilizing its resources. Thus, EDM decides to remove some existing instances ensuring the minimal QoE required and the load range of [$l^{min}$, $l^{max}$]. *Algorithm 1* describes more precisely the EDM functionality for an elastic allocation and re-allocation of virtual resources for a running CDN slice consisting of multiple VNFs. To summarize the main features of the algorithm in two main decisions, (i) EDM scales up and down the instances in terms of resources when the optimal solution is found; otherwise, (ii) it scales in and out the slice in case the LP model is infeasible.

---

6. In our implementation, we import the clouds and their flavors from a json file. For each cloud, we know the number of available flavors. Then we make sure that the vector $F = \{f_1, \ldots, f_B\}$ respects exactly the order of flavors retrieved from an ordered set of clouds. Thus, $k_{init}$ is considered as the index of the first flavor of the cloud $\zeta_k$, and $k_{fin}$ represents the index of the last flavor in that cloud.

---

**Algorithm 1** Elasticity Decision Maker Algorithm

---

**Require:** $\vartheta$ a set of running instances. $\xi$ a set of available cloud domains. $F$ a set of flavors. In a time period $T$.
  **while** Time period T **do**
    Receive QoE information from QA.
    Receive monitoring and resource usage from RUM.
    Run linear programming model (3): find an optimal allocation of flavors to running VNF instances.
    **while** Model is infeasible **do**
      Remove constraint (C3).
      Check if an optimum is found with a $load \geq l^{max}$.
      **if** Yes: possible optimization with over-utilization resources **then**
        Scaling out actions: add new VNF instance.
        Repeat the algorithm considering the newly added VNF instance.
      **end if**
      Restore constraint (C3).
      Remove constraint (C4).
      Check if an optimum is found with a $load \leq l^{min}$.
      **if** Yes: possible optimization with under-utilization resources **then**
        Scaling in actions: remove an existing VNF instance.
        Repeat the algorithm considering the newly removed VNF instance.
      **end if**
    **end while**
    Scaling up/down actions: Contact VIM(s) for new resource re-allocation to existing instances.
  **end while**

---

## 6.3 Efficiency and scalability of the EDM algorithm

### 6.3.1 Experiment setup

To evaluate our proposed solution, we implemented the EDM algorithm in Python. The linear programming model (3) is implemented using the Gurobi Optimization tool and is evaluated using the following metrics: (*i*) Mean cost per unit representing the average price paid in dollars per instance for its virtual resource allocation; (*ii*) QoE-aware resource allocation representing the performance when users require a minimal QoE to be ensured; (*iii*) execution time, representing the time required for the EDM algorithm to take an optimal decision. The optimization problem is solved for varying numbers of VNF instances in a CDN slice and other parameters related to the slice owner or user requirements, including the open sessions load and the minimum QoE required. In our simulations, we consider 45 cloud domains and 1,417 flavors in total with their respective pricing in $/hour. This information is collected from real IaaS platforms such as AWS service, Microsoft Azure and Rackspace. Hence, using real data reflects accurate and meaningful findings in our final simulation results. For the sake of simplicity, the total number of available CPU cores $C^{max}$ in each cloud domain is set to 20, and the impact of the flavor cost on QoE of a single instance is set to $\sigma = 0.1$. Finally, we run the simulations varying the number of VNF instances from 10 to 300 with a step of 10. Every instance is defined by an initial flavor selected uniformly at random from the set of 1,417 flavors, and a number of active parallel streaming sessions chosen such that the MOS estimated by 2 is non-negative. In our simulation results (i.e., Figs. 6 and 7), each plotted point represents the average of 35 executions.

### 6.3.2 *Performance with QoE-awareness & load-awareness*

Fig. 6.a shows the performance of our EDM algorithm in terms of mean cost of flavor per a deployed VNF instance. In this experiment, we set the average session load in a CDN slice to 0.7, and run the EDM algorithm to find the optimal assignment $x_{ij}^{opt}$ of flavors to VNF instances with the objective of minimizing the total deployment cost. Fig. 6.a shows that the mean cost per unit is much reduced when a lower QoE is required. For a QoE exceeding 3.5, the deployment of 200 VNF instances will cost 24.1$ per month for each instance. However, it will cost only 15.4$ per month in case the minimal QoE required is 2.5. Additionally, Fig. 6.b shows an intuitive result that the mean cost per unit is much higher in case of an over-loaded system. For a system load greater than 50%, the deployment of 200 VNF instances will cost 71.7$ per month for each instance. However, it will cost nearly 142.2$ per month in case the minimal load required is 80%. Figs. 6.a and 6.b also show an increasing mean cost per unit when increasing the size of the CDN slice. This is due to the limitations at the hosting cloud domain in terms of the amount of available resources and the maximum session load an instance can handle.

Based on the obtained results, the EDM algorithm is able to take reverse decisions. When the end-users require a low QoE and the slice owner has a budget limitation, the EDM algorithm can reversely propose the maximal number of VNF instances that can be deployed in a CDN slice respecting the owner's specified budget.

### 6.3.3 *Scalability of the EDM algorithm*

Fig. 6.c presents the execution times of the proposed EDM algorithm for different sizes of the target CDN slice. As shown, for a fixed number of flavors, execution time is roughly linear to the number of VNF instances. The execution time increased from 1.5 s in a CDN slice of 30 instances to 16 s for a slice of 300 instances. Hence, for large slices, our proposed EDM algorithm does not exceed 20 s to find the optimal resource allocation out of the available 1,417 flavors. We got similar results when changing other parameters, including the minimum quality of experience required by users in a CDN slice ($Q^{min}$), the minimum and maximum allowed loads in VNF instance ($l^{min}$, $l^{max}$), the maximal capacity in cloud infrastructures ($C^{max}$) and other key parameters.

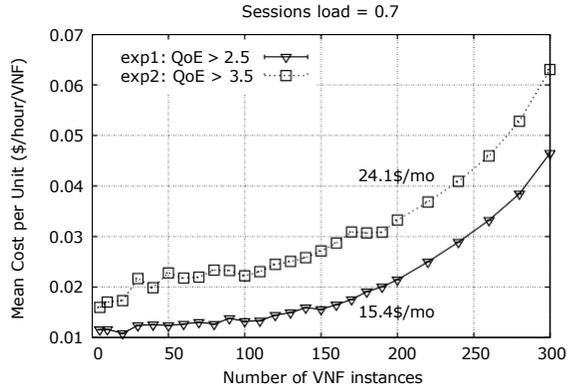

(a) Mean cost per instance in case of different QoE requirements and for varying numbers of VNF instances. The minimum session load is set to 0.7.

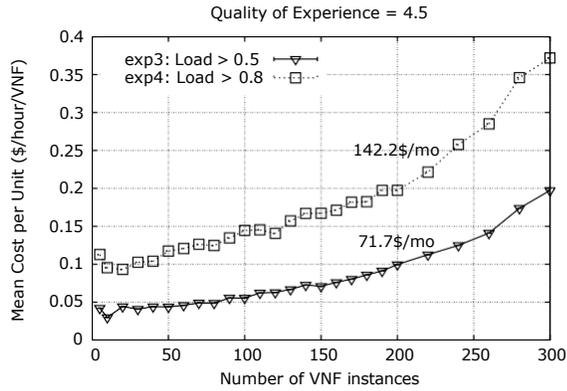

(b) Mean cost per instance in case of different session loads and for varying numbers of VNF instances. The minimum QoE is set to 4.5.

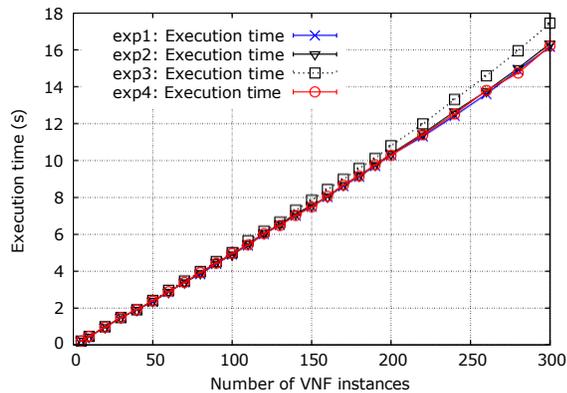

(c) The EDM algorithm's execution times in seconds. The same results obtained when changing values of other system parameters.

Fig. 6. Mean cost per unit and the EDM algorithm's execution time as a function of the CDN slice size. The algorithm is running on a KVM VM with 16 CPU cores.

### 6.3.4 *QoE behavior against system load*

Fig. 7 presents the average QoE experienced by end-users of a CDN slice of different sizes and that is under different loads. Fig. 7.a shows the correlation between load and QoE when the minimum system load and the minimum desired QoE are set to 50% and 4, respectively. Fig. 7.b shows the same in case of a higher system load (greater than 70%) and a higher QoE (greater than 4.5). Both Fig. 7.a and Fig. 7.b show that with less system load, the CDN slice performs better and exhibits better QoE values, exceeding always the minimum QoE value.

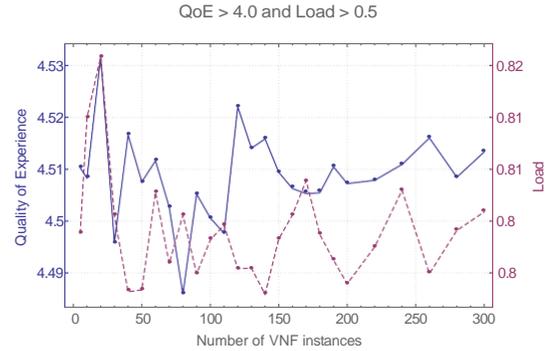

(a) Average QoE behavior vs. average load variation, when minimizing the deployment pricing cost with minimum requirements: minimal QoE = 4 and minimal load = 0.5.

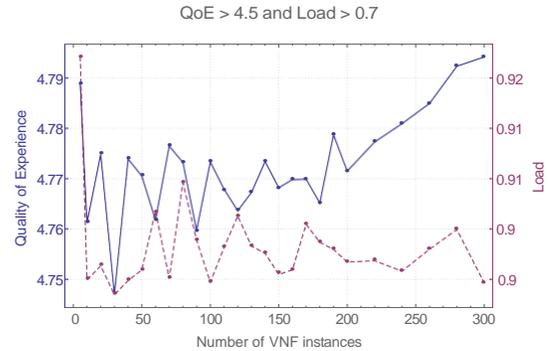

(b) Average QoE behavior vs. average load variation, when minimizing the deployment pricing cost with minimum requirements: minimal QoE = 4.5 and minimal load = 0.7.

Fig. 7. Average QoE behavior vs average load variation in a CDN slice when increasing the number of VNF instances.

## 7 ON-THE-FLY TRANSCODING

In case quality degradation is detected for a user based on significant changes in MOS values, and as soon as the QA identifies that this quality degradation is *not* due to increased compute load on the VNF instance that serves the user, it triggers the initiation of a video transcoding service. In particular, QA notifies the CDN-O, which in turn triggers the MEO to instantiate the appropriate functionality on the edge cloud. This service can also be triggered when the available network capacity in the RAN decreases or the cell load increases. This information is available via the RUM and can be exploited as described in Section 5.1.

Each slice contains an edge transcoding service deployed as a MEC application by the MEO upon request from the CDN-O (see Fig. 1). It is used to transcode the source media format to another deliverable format with a different bit-rate. Once the required operation is finished and the target QoE level is achieved, the virtual transcoder instance can be terminated and removed from the edge node to release resources.

For the purpose of demonstration, let us consider that the streaming server stores HLS content with a bit-rate of R1 (high bit-rate). As explained earlier, we assume that client-side functionality for recording and communicating to the QA playout status information is available. To start accessing the video service, a user connects to the edge and sends an HTTP GET request to the server to fetch the media description file (playlist or manifest file). Upon receiving the file, the end-user device starts sending HTTP GET requests to fetch the media segments sequentially as specified in the media description file. Once the initial necessary playout buffer level is achieved, the player starts displaying the video. At the end of every segment played, the reporter sends QoE-related information about the segment's playout to the QA. Based on this input, the QA periodically calculates a QoE estimate (MOS value). If the estimated QoE value is above the optimal, no further action is required. On the other hand, if the value decreases, the QA signals the CDN-O to spin up the transcoding function.

The virtual transcoder instance is launched by MEO using an image stored at the edge cloud's image store, as specified in the CDNaaS slice description template. The video bit-rate is reduced to R2 ($R2 < R1$), one step lower than the current bit-rate. When the transcoder signals the end of its task, a mixer service is used which uses information on the last downloaded segment ID to replace the segments not yet fetched by the client with the newly encoded ones. Once the operation is finished and if MEO does not receive any other transcoding requests for a given time interval, it terminates the virtual transcoder instance (container or VM). However, if the QoE estimated by QA remains low, the same operation is triggered again, reducing the video bit-rate to R3 ($R3 < R2$). This stepwise recurring operation continues until a target optimal QoE is achieved at the client end.[7]

This transcoding approach has some advantages. Depending on the storage capacity at the edge, storage overhead can be reduced, as pre-transcoded multiple versions of the same content are not required from the beginning and are generated on-demand. On the other hand, if storage capacity is adequate, the original and the transcoded video chunks can be maintained by the cache VNF instance, and reused in the future, thus avoiding the computationally intensive transcoding process for future requests for the same video content. Moreover, by serving the content locally (i.e., from the edge), the solution ensures reduction in core network traffic and reduced end-to-end latency.

Given these last points, further enhancements are possible for the virtual transcoding service. Considering a mobile network, where network conditions are dynamically changing due to user mobility, high sensitivity towards transient events may lead to a "ping pong" effect, where transcoding will be initiated every now and then following the variations in the conditions of the network. Instead, transcoding should be triggered only if the network conditions have actually degraded. However, still the tradeoff between responsiveness and avoiding this effect exists. Maintaining a sliding window of QoE scores to decide on whether to initiate transcoding based on a running average of the MOS (within that window) can help address this issue.

Various further performance enhancements are also possible. For example, the cross-layer mobility, bandwidth [58], and QoE [59] prediction mechanisms that we have studied in our prior work in similar contexts can be applied to this end. Such an ability to predict the conditions at the client end can assist in identifying the optimal time to initiate transcoding. Depending on the resource availability in the edge host, the transcoding service can be initiated in parallel to support multiple end-user requests. Considering it as a compute-intensive task, and in case of limited resources in the serving edge, it is the MEO's responsibility to select another nearby edge host (taking into account response time and resource availability) to perform the transcoding-only operation, building on the shared infrastructure concept of MEC. Edge host selection algorithms are outside the scope of this work, but also of that of the ETSI MEC standard.

## 8 TESTBED EXPERIMENTS

We carry out two sets of testbed experiments to evaluate the performance of our dynamic slice resource management mechanisms. First, we focus on the performance of our QoE- and load-aware elastic compute resource management scheme. We then present experimental results on the use of the virtual transcoding service.

### 8.1 Elastic cloud resource management

To evaluate our cloud resource management scheme, we set up a testbed using a Ubuntu 14.04.03 LTS desktop workstation with 8 CPU cores and 16 GB RAM. The cloud environment was built using Openstack (Devstack Juno version) inside a VirtualBox Ubuntu server with dedicated 4 vCPUs and 8 GB RAM. The setup consists of an all-in-one Openstack environment with the controller, compute, heat and neutron components running on the same node. Heat was used to orchestrate the initial setup of a round-robin load balancer (LB) and a single instance within the LB pool with a built-in NGINX server for streaming a preloaded video file. HTTP-based live network streaming (i.e., progressive streaming) was used. For load generation we used ApacheBench and a modified version of the *wrk* tool, while we used the VLC media player with our modifications to record interruptions and video quality parameters (the same software we used in the experiments of Section 4), so that we can directly calculate QoE (MOS) at the user end.

VNF instances were created from a Ubuntu cloud image using a customized flavor with either 1 vCPU and 512 MB RAM (Flavor 1) or 2 vCPUs and 2 GB RAM (Flavor 2). The RUM, QA and EDM modules were executed in the same node as the cloud controller. At the RUM, resource usage information is monitored and fed into the EDM every 60 s. To introduce high load in the VM hosting the video service (i.e., busy CPU and exhausted RAM), 2000 concurrent connections (a total of 100000 HTTP requests/s) were launched towards the server.

Following our experiments of § 4, as Fig. 2 and Fig. 3 indicate, when CPU load approaches full utilization, interruptions start to take place and QoE drops. Therefore, we set resource usage thresholds to 90% for both CPU and RAM as the upper limit to trigger a scale up operation. As for scaling down, the thresholds were set to 25% and 65% for CPU and RAM, respectively. These thresholds were selected also with the intention to keep the system responsiveness in enforcing elasticity (i.e., particularly scaling up) within a reasonable range; if the system takes too long to respond, the service will be totally disrupted before even scaling up the relevant instance.

---

[7]. To speed up the convergence of this procedure, the initial value for R2 can be selected considering an estimate of the maximum attainable throughput as a function of the user's CQI value reported by the RNIS, as described in Section 5.1. The stepwise procedure might still have to be carried out, since the initial estimate might not be fully accurate.

To demonstrate elastic CDN slice resource management, a VNF instance is created using Flavor 1 and once its resource usage reaches the upper thresholds, EDM triggers Heat (i.e., the Openstack orchestration engine implementing some CDNaaS service orchestration functionality in our tests) to launch a new instance using Flavor 2 and assign it to the same load balancer pool. Once the second instance becomes active, the first one is released ensuring the service is uninterrupted. Scaling down was also carried out in a similar fashion. Notably, since the instances were active, resize or update operations were not performed, as they would suspend the image, thus disrupting the video streaming service.

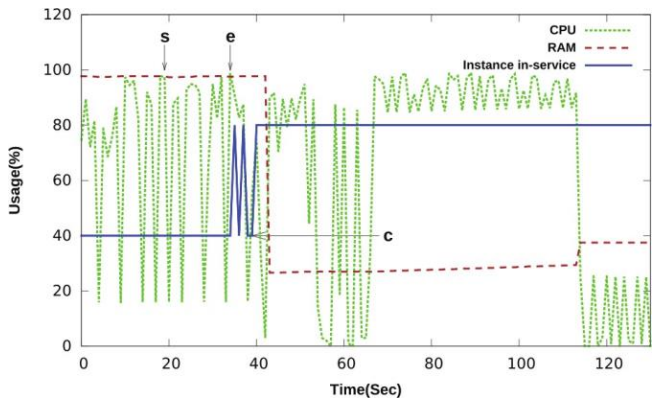

Fig. 8. An example of a scale up operation. (s: elasticity trigger time; e: VM migration end time; c: VM actual changeover time).

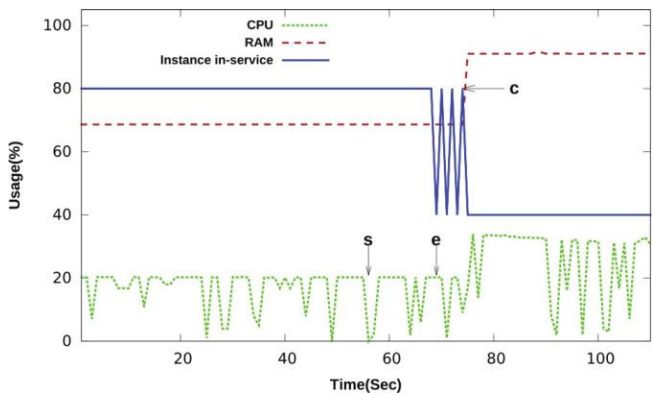

Fig. 9. An example of a scale down operation (s: elasticity trigger time; e: VM migration end time; c: VM actual changeover time).

Fig. 8 and Fig. 9 depict the full scale-up and scale-down operations, respectively. The solid line indicates which instance is in service and how the load balancer takes care of the switch-over. The solid line at the 80% level indicates that Flavor 2 is used for streaming and at the 40% level indicates that Flavor 1 is in use. The pointers 's', 'e', and 'c' on Fig. 8 and Fig. 9 indicate respectively the elasticity triggering start time, the end of the scaling operation, and the actual instance changeover time considering a predefined sleep time set up for the new instance to ensure service continuity.

These figures show that right after the flavor changeover, the usage ratios of both RAM and CPU change, although the injected load is constant in all cases. The figures also indicate that the LB distributes the service between the two instances as soon as the new instance is created. However, the complete release of the old instance takes some time as a sleep time interval was deliberately injected to ensure a smooth streaming service. It is worth noting that the delay since triggering vertical scaling until the release of old instance is merely 15 to 20 s.

## 8.2 Virtual transcoding service

In this section, we describe our testbed environment which emulates a CDN slice with streaming and on-demand transcoding components at the mobile edge. We assume that the media content for adaptive streaming is already stored at the MEC streaming server. The experimental environment mostly focuses on assessing the performance of the edge, which is emulated by two laptops (running Ubuntu 14.04.3 LTS); one represents the edge node and the other the client.

At the client end, we used our modified VLC video player. Relevant playout statistics (i.e., interruptions and quantization parameters) were retrieved by the reporter component (implemented in python), filtered, and passed on to the QA, which, in our tests, was running on the edge node for simplicity. At the edge side, two VirtualBox VMs were used. VM1 was used as a gateway for the entire network to access the Internet. DHCP with authentication was also set up inside VM1 to configure the whole network using a single subnet (to ease the emulation). VM2 was configured using the Proxmox Virtual Environment to act as the Mobile Edge Host. We use OpenVZ containers to host the CDN slice components as ME Applications at the edge host. The streaming server functionality was provided by a Ubuntu cloud minimal image with the NGINX webserver installed, instantiated as an Openvz container. NGINX was configured for HLS streaming. The content for initial streaming was pre-transcoded and prepared using ffmpeg, and the video codec used was H.264/AVC. The same container is used for the storage of the media files. The QA was configured to receive interruption and QP information periodically from the reporter and appropriately normalize/transform them to be used as input to the PSQA tool to calculate the expected MOS values. The same QA module was responsible for the evaluation of the MOS values and for triggering the component which was emulating the MEO (also implemented in Python). The MEO was responsible for spinning up the container which provides the transcoding service. The transcoding container template was based on a Ubuntu minimal image and had ffmpeg installed.

Our setup also includes a mixer service responsible for replacing the original content with the transcoded one in the streaming server (this functionality was also built into the same container). Finally, to ensure that the laptop acted as an edge access point, its Wi-Fi interface was configured using hostapd in IEEE 802.11 master mode. The netem and wondershaper tools were used to simulate a cellular network environment.

One desired feature of the on-the-fly transcoding service is responsiveness, which we define as the measured delay from the time of triggering the service to the actual QoE enhancement time. This responsiveness check was performed in two ways:

- Case (a): The container is already active with pre-transcoded media files. Only the mixer functionality is used, and, thus, taken into account in the delay measurement.
- Case (b): Using the full functionality of on-the-fly transcoding by triggering the creation of the transcoder virtual instance as a ME application, initiating the service, transcoding the media file and then performing the mixing.

The media file used for this test was 9m56s long with 298 segments in total, each carrying 2 s of video. The reporter information was sent after 2 s of video had been played. To perform the QA operation, the tool requires the information of 16 s of played video. Therefore, the MOS calculation was performed only after information of a total of 8 segments were received. A MOS value below 3.5 was considered low and was used to trigger the MEO.

Case (a) is represented in Fig. 10, where initially the MOS value was high. With time, it degraded and as soon as it reached below the predefined threshold, mixing was initiated. In this scenario, pre-transcoded low bit-rate files were copied to the streaming server's and replaced the existing ones. To save time, only the segments which were yet to be downloaded by the client were replaced. The replacement occurs in a sequential manner. As a result, although the mixing time was approximately 26 s, QoE started enhancing as soon as few segments were delivered.

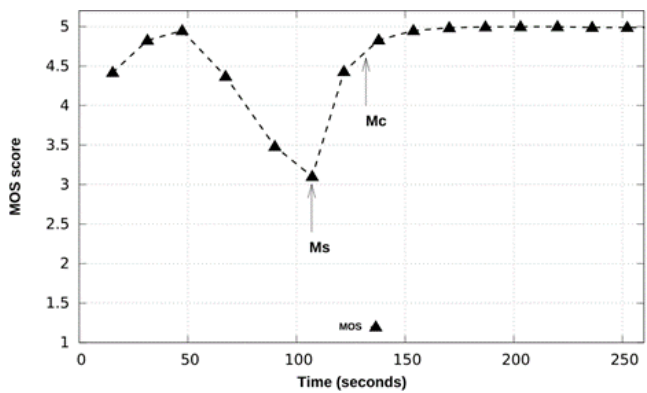

Fig. 10. MOS vs. Time (with pre-transcoded media files). (Ms: mixing start time; Mc: mixing completion time).

Fig. 11 depicts the full functionality as mentioned in case (b). "Ti" represents the time when container instantiation started. The time difference between the QA triggering the MEO and the MEO starting the creation of the container is in the ms range. The container boots up with the already prepared transcoder image template (within 3 s). Once it is ready, the MEO signals to start the transcoding with the specified rate. Having limited resources (2 vCPUs and 1024 MB RAM), the container performs this total operation in approximately 70 s. The mixer operation starts as soon as transcoding is over. In between, the mixer receives the information on the last downloaded file to identify from which segment mixing will start. The mixer operation takes almost similar time as mentioned before (in case (a)) and depends on the number of segments to be transferred and replaced.

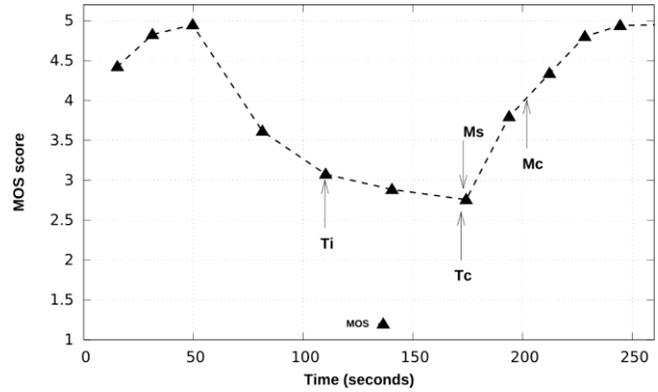

Fig. 11. MOS vs. Time (full functionality). (Ti: transcoding container initiated; Tc: transcoding completed; Ms: mixing started; Mc: mixing completed).

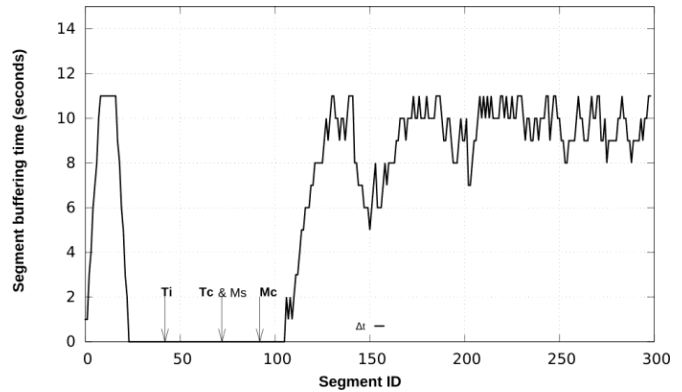

Fig. 12. Segment buffering time ($\Delta t$) vs Segment ID. (Ti: transcoding container initiated; Tc: transcoding completed; Ms: mixing started; Mc: mixing completed; when segment buffering time = '0', corresponding segments are not downloaded in advance).

In Fig. 12, the segment buffering time ($\Delta t$) is the time a downloaded segment spends in the buffer waiting for playout. In other words, it is the time difference between the instant its playout started and the instant it was fully downloaded. It is clear that initially the segment buffering time was high, which indicates that the segments were downloaded in advance. As a result, the immediate next few segments were already downloaded and ready before it was being played. Therefore, there was no buffering delay, and thus no playout interruptions, hence the MOS was high (if compared with Fig. 11). When we introduced a degradation in network conditions, emulating congestion due to significant background traffic, the downloading segment time in-creased because of a reduction in the available bandwidth. At a point when the difference ($\Delta t$) was almost zero, the video experienced buffering delays, as it had to wait until the download of a segment is complete. However, after transcoding to a lower bit-rate video, the segment size, as well as the segment download time, reduced. Consequently, adapting video bandwidth demands to the current network conditions led to timely video segment downloads, and minimized playback buffering time. This positively impacted the overall MOS.

From Fig. 11, it is evident that with the degradation in network conditions, the MOS value started reducing and at a certain point it reached 3 (below the acceptable QoE threshold). In such a situation, if on-the-fly transcoding were not enforced, the CDN slice end user would have experienced reduced video quality for the entire remaining duration (486 s). However, instantiating a transcoder at the edge enabled the client to experience low quality only for a limited amount of time (from detecting MOS degradation until the end of transcoding and mixing, i.e., 96 s), after which the viewing experience of the remaining video (390 s) improved.

## 9 CONCLUSION

NFV and MEC open new possibilities for innovative services for the forthcoming 5G mobile communications era. In this article, taking advantage of these specific technologies, we expanded on the architectural and technical aspects of the dynamic instantiation and elastic management of CDN slices over shared, multi-domain, edge (and other) cloud infrastructures. The design and mechanisms we proposed come with improved QoE awareness. This is valuable both at service instantiation time, when our empirical models facilitate informed CDN slice resource dimensioning decisions, but also at run time, when multi-level monitoring information are applied to resource scaling and video-service-level adaptations, in order to optimize service quality and resource utilization. NFV and MEC standards, and the 5G architecture in general, continue to evolve. Our ongoing and future work will focus on adapting our design following these developments, at the same time working on service-level optimizations. These include more sophisticated resource scaling mechanisms, optimized VNF placement algorithms, and more efficient video adaptation schemes. Aspects related with SLA design and monitoring, as well as other business-relevant issues including service pricing are of equal importance for further study.

## ACKNOWLEDGMENT

This work was partially funded by the Academy of Finland Project CSN under Grant Agreement No. 311654, and the Academy of Finland's Flagship Programme 6Genesis under Grant Agreement No. 318927. The work is also partially supported by the European Union's Horizon 2020 research and innovation program under the 5G!Pagoda, 5G-TRANSFORMER, and MATILDA projects with grant agreement No. 723172, No. 761536, and No. 761898, respectively.

## REFERENCES

[1] Cisco, "Cisco Visual Networking Index: Global Mobile Data Traffic Forecast Update, 2015–2020," Cisco, White Paper, Feb. 2016.
[2] I. Benkacem, T. Taleb, M. Bagaa, and H. Flinck, "Optimal VNFs placement in CDN Slicing over Multi-Cloud Environment," *IEEE J. Sel. Areas Commun.*, vol. 36, no. 3, Mar. 2018.
[3] M. Patel et al., "Mobile-Edge Computing – Introductory Technical White Paper," ETSI, White Paper, September 2015.
[4] Y. C. Hu, M. Patel, D. Sabella, N. Sprecher, and V. Young, "Mobile Edge Computing–A key technology towards 5G," ETSI, White Paper 11, Sep. 2015.
[5] T. Taleb, K. Samdanis, B. Mada, H. Flinck, S. Dutta, and D. Sabella, "On Multi-Access Edge Computing: A Survey of the Emerging 5G Network Edge Cloud Architecture and Orchestration," *IEEE Commun. Surveys Tuts.*, vol. 19, no. 3, pp. 1657–1681, 2017.
[6] *Mobile Edge Computing (MEC); Framework and Reference Architecture*, ETSI Group Specification MEC 003, Oct. 2017.
[7] *Network Functions Virtualisation (NFV); Management and Orchestration*, ETSI Group Specification NFV-MAN 001, Dec. 2014.
[8] R. Jain and S. Paul, "Network virtualization and software defined networking for cloud computing: a survey," *IEEE Commun. Mag.*, vol. 51, no. 11, pp. 24–31, 2013.
[9] *Network Functions Virtualisation (NFV); Use Cases*, ETSI Group Specification NFV 001, Oct. 2013.
[10] FP7 T-NOVA. [Online]. Available: http://www.t-nova.eu
[11] G. Xilouris *et al.*, "T-NOVA: A marketplace for virtualized network functions," in *Proc. EuCNC*, 2014.
[12] H. Chang, A. Hari, S. Mukherjee, and T. Lakshman, "Bringing the cloud to the edge," in *Proc. IEEE INFOCOM Workshops*, 2014.
[13] T. Taleb, S. Dutta, A. Ksentini, I. Muddesar, and H. Flinck, "Mobile Edge Computing Potential in Making Cities Smarter," *IEEE Commun. Mag.*, vol. 55, no. 3, pp. 38–43, 2017.
[14] V. Sciancalepore, F. Giust, K. Samdanis, and Z. Yousaf, "A double-tier MEC-NFV architecture: Design and optimisation," in *Proc. IEEE CSCN*, Nov. 2016.
[15] D. Fesehaye, Y. Gao, K. Nahrstedt, and G. Wang, "Impact of Cloudlets on Interactive Mobile Cloud Applications," in *Proc. IEEE 16th Int'l Conf. on Enterprise Distributed Object Computing Conference (EDOC)*, Sep. 2012.
[16] J. Fajardo, I. Taboada, and F. Liberal, "Improving content delivery efficiency through multi-layer mobile edge adaptation," *IEEE Network*, vol. 29, no. 6, pp. 40–46, 2015.
[17] W. Zhu, C. Luo, J. Wang, and S. Li, "Multimedia Cloud Computing," *IEEE Signal Process. Mag.*, vol. 28, no. 3, pp. 59–69, May 2011.
[18] Y. Jararweh, L.Tawalbeh, F. Ababneh, and F.Dosari, "Resource Efficient Mobile Computing Using Cloudlet Infrastructure," in *Proc. IEEE MSN*, Dec. 2013.
[19] S. Dutta, T. Taleb, P. A. Frangoudis, and A. Ksentini, "On-the-Fly QoE-Aware Transcoding in the Mobile Edge," in *Proc. IEEE Globecom*, 2016.
[20] L. Wei, J. Cai, C. H. Foh, and B. He, "Qos-aware resource allocation for video transcoding in clouds," *IEEE Trans. Circuits Syst. Video Technol.*, vol. 27, no. 1, pp. 49–61, Jan 2017.
[21] X. Li, M. A. Salehi, M. Bayoumi, N. Tzeng, and R. Buyya, "Cost-efficient and robust on-demand video transcoding using heterogeneous cloud services," *IEEE Trans. Parallel Distrib. Syst.*, vol. 29, no. 3, pp. 556–571, March 2018.
[22] F. Wang, J. Liu, and M. Chen, "CALMS: Cloud-Assisted Live Media Streaming for Globalized Demands with Time/Region Diversities," in *Proc. IEEE INFOCOM*, Mar. 2012.
[23] Y. Zha, H. Jiang, K. Zhou, Z. Huang, and P. Huang, "Meeting Service Level Agreement Cost-Effectively for Video-on-Demand Applications in the Cloud," in *Proc. IEEE INFOCOM*, May 2014.
[24] R. Cheng, W. Wu, Y. Lou, and Y. Chen, "A Cloud-Based Transcoding Framework for Real-Time Mobile Video Conferencing System," in *Proc. IEEE MobileCloud*, Apr. 2014.
[25] Y. Wu, C. Wu, B. Li, and F. C. Lau, "vSkyConf: Cloud-assisted multiparty mobile video conferencing," in *Proc. 2nd ACM SIGCOMM Workshop on Mobile Cloud Computing*, Aug. 2013.
[26] M. Chen, "AMVSC: A framework of adaptive mobile video streaming in the cloud," in *Proc. IEEE Globecom*, Dec. 2013.
[27] W. Zhang, Y. Wen, and H.-H. Chen, "Toward Transcoding as a Service: Energy-Efficient Offloading Policy for Green Mobile Cloud," *IEEE Network*, vol. 28, no. 6, pp. 67–73, Nov. 2014.
[28] F. Jokhio, T. Deneke, S. Lafond, and J. Lilius, "Bit rate reduction video transcoding with distributed computing," in *Proc. PDP*, Feb. 2012.

[29] F. Lao, X. Zhang, and Z. Guo, "Parallelizing video transcoding using map-reduce-based cloud computing," in *Proc. IEEE ISCAS*, May 2012.
[30] G. Gao, W. Zhang, Y. Wen, Z. Wang, W. Zhu, and Y. P. Tan, "Cost optimal video transcoding in media cloud: Insights from user viewing pattern," in *Proc. IEEE ICME*, Jul. 2014.
[31] A. Heikkinen, J. Sarvanko, M. Rautiainen, and M. Ylianttila, "Distributed multimedia content analysis with mapreduce," in *Proc. IEEE PIMRC*, Sep. 2013.
[32] M. Kim, S. Han, Y. Cui, H. Lee, H. Cho, and S. Hwang, "Cloud-DMSS: Robust Hadoop-Based multimedia streaming service architecture for a cloud computing environment," *Cluster Computing*, vol. 17, no. 3, pp. 605–628, Sep. 2014.
[33] Amazon elastic transcoder. [Online]. Available: https://aws.amazon.com/elastictranscoder/
[34] M. T. Beck, S. Feld, A. Fichtner, C. Linnhoff-Popien, and T. Schimper, "ME-VoLTE: Network functions for energy-efficient video transcoding at the mobile edge," in *Proc. ICIN*, Feb. 2015.
[35] Ericsson, "5G System: Enabling Industry and Society Transformation," Ericsson, White Paper 284 23-3251 Uen, Jan. 2015.
[36] Architecture Working Group, "View on 5G Architecture," 5G PPP, White Paper, Jul. 2016.
[37] I. Afolabi, T. Taleb, K. Samdanis, A. Ksentini, and H. Flinck, "Network Slicing and Softwarization: A Survey on Principles, Enabling Technologies and Solutions," *IEEE Commun. Surveys Tuts.*, vol. 20, no. 3, pp. 2429–2453, 2018.
[38] NGMN P1 WS1 E2E Architecture Team, "Description of network slicing concept," NGMN Alliance, Final Deliverable, Jan. 2016.
[39] Q. Li, G. Wu, A. Papathanassiou, and M. Udayan, "An end-to-end network slicing framework for 5G wireless communication systems," *CoRR*, vol. abs/1608.00572, 2016.
[40] N. Nikaein *et al.*, "Network store: Exploring slicing in future 5G networks," in *Proc. ACM MobiArch*, 2015.
[41] T. Taleb, A. Ksentini, and R. Jantti, ""Anything as a Service" for 5G Mobile Systems," *IEEE Network*, vol. 30, no. 6, pp. 84–91, Nov. 2016.
[42] P. A. Frangoudis, L. Yala, and A. Ksentini, "CDN-as-a-Service provision over a telecom operator's cloud," *IEEE Trans. Netw. Service Manag.*, vol. 14, no. 3, pp. 702–716, 2017.
[43] S. Dutta, T. Taleb, and A. Ksentini, "QoE-aware Elasticity Support in Cloud-Native 5G Systems," in *Proc. IEEE ICC*, May 2016.
[44] *Mobile Edge Computing (MEC); Radio Network Information API*, ETSI Group Specification MEC 012, Jul. 2017.
[45] *Mobile Edge Computing (MEC); Mobile Edge Management; Part 2: Application lifecycle, rules and requirements management*, ETSI Group Specification MEC 010-2, Jul. 2017.
[46] *Mobile Edge Computing (MEC); Deployment of Mobile Edge Computing in an NFV environment*, ETSI Group Report MEC 017, Feb. 2018.
[47] *Multi-access Edge Computing (MEC); MEC support for network slicing*, ETSI Work Item MEC 024.
[48] S. Kekki *et al.*, "MEC in 5G Networks," ETSI, White Paper 28, Jun. 2018.
[49] *Information technology – Dynamic adaptive streaming over HTTP (DASH) – Part 1: Media presentation description and segment formats*, ISO/IEC Standard 23 009-1:2014, May 2014.
[50] nginx. [Online]. Available: http://nginx.org
[51] A. Kivity, Y. Kamay, D. Laor, U. Lublin, and A. Liguori, "kvm: the Linux virtual machine monitor," in *Proc. Linux Symposium*, 2007.
[52] Docker. [Online]. Available: https://www.docker.com/
[53] G. Rubino, "Quantifying the Quality of Audio and Video Transmissions over the Internet: The PSQA Approach," in *Communication Networks & Computer Systems*, J. A. Barria, Ed. Imperial College Press, 2005.
[54] K. D. Singh, Y. H. Aoul, and G. Rubino, "Quality of experience estimation for adaptive HTTP/TCP video streaming using H.264/AVC," in *Proc. IEEE CCNC*, 2012.
[55] S. Arora, P. A. Frangoudis, and A. Ksentini, "Exposing radio network information in a MEC-in-NFV environment: the RNISaaS concept," EURECOM, Research Report RR-19-339, Feb. 2019.
[56] *LTE; Evolved Universal Terrestrial Radio Access (E-UTRA); Physical layer procedures (3GPP TS 36.213 version 14.2.0 Release 14)*, ETSI Technical Specification ETSI TS 136 213, Apr. 2017.
[57] DASH Industry Forum Software. [Online]. Available: http://dashif.org/software/
[58] A. Nadembega, A. Hafid, and T. Taleb, "An integrated predictive mobile-oriented bandwidth-reservation framework to support mobile multimedia streaming," *IEEE Trans. Wireless Commun.*, vol. 13, no. 12, pp. 6863–6875, Dec. 2014.
[59] A. Ksentini, T. Taleb, and K. Letaif, "QoE-Based Flow Admission Control in Small Cell Networks," *IEEE Trans. Wireless Commun.*, vol. 15, no. 4, pp. 2474–2483, Apr. 2016.

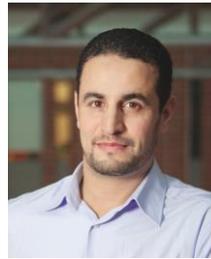
**Tarik Taleb** is a professor at Aalto University. He is the founder and director of the MOSA!C Lab (www.mosaic-lab.org). Prior to that, he was a senior researcher and 3GPP standards expert at NEC Europe Ltd., Germany. He also worked as an assistant professor at Tohoku University, Japan. He received his B.E. degree in information engineering, and his M.Sc. and Ph.D. degrees in information sciences from Tohoku University in 2001, 2003, and 2005, respectively.

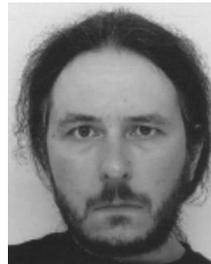
**Pantelis A. Frangoudis** received the B.Sc. (2003), M.Sc. (2005), and Ph.D. (2012) degrees in CS from the Department of Informatics, AUEB, Greece. He has been a post-doctoral researcher (2012-2017) with Team Dionysos, IRISA/INRIA/University of Rennes 1, Rennes, France. Currently, he is a researcher with the Communication Systems Department, EURECOM, Sophia Antipolis, France.

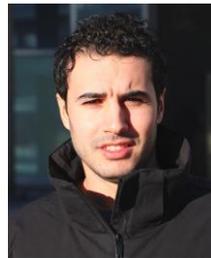
**Ilias Benkacem** received the B.Sc. degree in Mathematics and Physics Higher School Preparatory Classes for Engineering Schools (CPGE), Tangier, Morocco, in 2013, and the Engineer's degree from the National Superior School of Computer Science and System Analysis (ensias), Mohammed V University, Rabat, Morocco, in 2016. He is currently pursuing the Ph.D. degree with the School of Electrical Engineering, Aalto University, Finland.

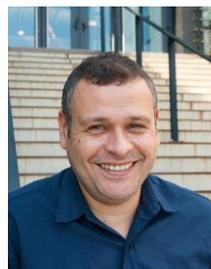
**Adlen Ksentini** is an IEEE Communications Society Distinguished Lecturer. He obtained his Ph.D. in computer science from the University of Cergy-Pontoise in 2005. From 2006 to 2016, he was an assistant professor at the University of Rennes 1. In March 2016, he joined the Communication Systems Department of EURECOM as an assistant professor. He has been working on network slicing in the context of the EU-funded H2020 projects 5G!Pagoda and 5G-TRANSFORMER.